\documentclass[reprint,amsmath,amssymb,aps,superscriptaddress,nofootinbib]{revtex4-2}
\usepackage{graphicx}
\usepackage{dcolumn}
\usepackage{bm}
\usepackage[utf8]{inputenc}
\usepackage{float}
\usepackage{hyperref}
\usepackage{amsmath, amsthm, amsfonts,amssymb}
\usepackage[svgnames]{xcolor}
\definecolor{teal}{RGB}{0, 128, 128}
\definecolor{myred}{RGB}{179, 27, 27}
\hypersetup{
    colorlinks=true,
    linkcolor=teal, 
    citecolor=teal, 
    urlcolor=teal  
 }
\usepackage{physics}
\usepackage{array}
\usepackage{booktabs}
\usepackage{dsfont}
\usepackage{graphicx}
\usepackage{cleveref}
\usepackage{xfrac}
\usepackage{mathrsfs}
\usepackage{comment}
\usepackage{adjustbox}

\def\apj{{ApJ}}

\def\aap{{A\&A}}

\def\araa{{Annual Review of Astronomy and Astrophysics}}

\def\mnras{{MNRAS}}

\def\nat{{Nature}}

\def\prd{{Physical Review D}}
\def\prl{{Phys. Rev. Lett.}}

\def\04a{{2004 a}}
\def\04b{{2004 b}}

\begin{document}
\title{Lensing by black holes within astrophysical environments}
\author{Gerasimos Kouniatalis}
\email{gkouniatalis@noa.gr}
\affiliation{Physics Department, National Technical University of Athens, 15780 Zografou Campus, Athens, Greece}
\affiliation{National Observatory of Athens, Lofos Nymfon, 11852 Athens, Greece}
\author{Arthur G. Suvorov}
\affiliation{Departament de Física Aplicada, Universitat d’Alacant, Ap. Correus 99, E-03080 Alacant, Spain}
\affiliation{Theoretical Astrophysics, Institute for Astronomy and Astrophysics, University of T\"{u}bingen, 72076 T\"{u}bingen, Germany}
\author{Kyriakos Destounis}
\affiliation{CENTRA, Departamento de Física, Instituto Superior Técnico -- IST, Universidade
de Lisboa -- UL, Avenida Rovisco Pais 1, 1049-001 Lisboa, Portugal}
%
%%%%%%%%%%%%%%%%
\begin{abstract}
Astrophysical black holes are likely to be surrounded by various forms of matter in the form of disks or halos. While a number of studies have examined the impact of an environment on the lensing of light or gravitational waves from cosmological sources, these have, thus far, been carried out in either a Newtonian or post-Newtonian framework where the environment is superimposed on the black-hole spacetime. By using an exact solution in general relativity describing a black hole embedded within a realistic halo of Hernquist matter distribution, we study deflection angles and image amplification in a fully relativistic setup. It is shown that large ``bumps'', that also arise at the Newtonian and post-Newtonian levels, track the transition scale set by the halo parameters that control the strong-lensing upturn and can significantly adjust the inferences made for either the source or lens in various contexts. As an application, we consider ``echoes'' of gravitational waves, sourced by astrophysical lenses rather than being intrinsic to the compact object that produces the signal.
\end{abstract}
%%%%%%%%%%%%%%%%

\maketitle

%%%%%%%%%%%%%%%%%%%%%%%%%%%%%%%%%%%%%%%%
\section{Introduction} \label{sec:intro}
%%%%%%%%%%%%%%%%%%%%%%%%%%%%%%%%%%%%%%%%

Black holes (BHs) are a fundamental prediction of general relativity (GR) \cite{Penrose:1964wq,chandra83}. The classical Schwarzschild and Kerr solutions uniquely describe these objects in vacuum \cite{heus96}, and their properties have been studied extensively in a variety of contexts. In reality, however, BHs are unlikely to be truly isolated, and in fact they are typically detected due to their influence on the surrounding astrophysical environment \cite{Semerak:1998wn,Doeleman:2009te,Contopoulos:2018zfu,Barausse:2014tra,Toubiana:2020drf,Sberna:2022qbn}. For instance, they can be surrounded by accretion discs \cite{Lemos:1993qp,Karas:2004rj,Semerak:2020kbp,Kotlarik:2022spo,Speri:2022upm,Duque:2024mfw,Yuan:2025fde,Spieksma:2025wex}, embedded in globular clusters \cite{Kulkarni:1993fr,Sigurdsson:1993zrm,Miller:2001ez,Pfahl:2005nt,Strader:2012wj,Morscher:2014doa}, or enshrouded by various forms of plasma \cite{Chou:1999qna,Cardoso:2020nst,Atamurotov:2021hoq,Zwick:2021dlg,Dima:2020rzg,Cannizzaro:2020uap,Cannizzaro:2023ltu,Spieksma:2023vwl,Alonso-Monsalve:2023jfq,Cannizzaro:2024yee} or clouds of ordinary or exotic material \cite{Herdeiro:2015gia,Cunha:2015yba,Berti:2019wnn,Choudhary:2020pxy,Cannizzaro:2023jle,Brito:2023pyl,Chen:2022kzv,Maselli:2020zgv,Maselli:2021men,Barsanti:2022vvl,Cannizzaro:2024hdg,Brito:2025ojt}. The BHs residing at the center of galaxies and active galactic nuclei (AGN) \cite{Vitral:2021,Shin:2012,EventHorizonTelescope:2019dse,EventHorizonTelescope:2022wkp} are surrounded by halos \cite{Oort:1940,Rubin:1970,Begeman:1991iy,Persic:1995ru,Corbelli:1999af,Freese:2008cz}, the bulk of which may be made up of dark matter \cite{Navarro:1995iw,Navarro:1996gj,Graham:2005xx,Prada:2005mx,Borriello:2000rv,clowe06,Duque:2023seg}. Such halo structures trace primordial perturbations generated during inflation \cite{Kouniatalis:2025orn,Germani:2017bcs, Ballesteros:2017fsr,Ezquiaga:2017fvi}. For individual BHs, compact binaries, or even supermassive objects in the center of galaxies, the environment may modulate photons or gravitational waves (GWs) in a number of ways depending on the scales of the halos involved \cite{Cardoso:2019rou,Leong:2023nuk}; from the inspiral-merger stage \cite{Destounis:2022obl,CanevaSantoro:2023aol,Morton:2023wxg,Aurrekoetxea:2024cqd,Cole:2022yzw,Ghosh:2025ban,Destounis:2025tjn} to post-merger ringdown \cite{Cheung:2021bol,Destounis:2023ruj,Rosato:2024arw}. Revisiting classical predictions but for non-vacuum BHs is therefore of astrophysical relevance. 

Most studies of how BH dynamics are altered by dark matter environments have treated the problem perturbatively, where a halo or disc is essentially put by hand by superimposing a PN potential onto the vacuum metric \cite{Macedo:2013qea,Cardoso:2019upw,Polcar:2022bwv,Vicente:2025gsg}. Despite the widespread use and successes of perturbative approaches (see, e.g., Refs. \cite{Katz:2021yft,Bertone:2024rxe,Polcar:2025yto,Dyson:2025dlj}), the growing demand for high-accuracy templates for GWs, so as to be compared with existing and upcoming data via matched filtering \cite{Chan:2024qmb}, demands a fully-relativistic construction, from first-principles, of BHs embedded in astrophysical environments. A recent advancement in this direction was made by \cite{Cardoso:2021wlq} using the ``Einstein cluster'' scheme \cite{Einstein:1939ms,Geralico:2012jt} where an exact, static, and spherically-symmetric BH is lodged in a dark-matter halo with a Hernquist mass-density distribution \cite{Hernquist:1990be}. In this case, the structure of the halo is intrinsically imprinted on the spacetime. Other self-consistent solutions have since been found using different density profiles \cite{Jusufi:2022jxu,Konoplya:2022hbl,Shen:2023erj,Figueiredo:2023gas,Speeney:2024mas,Pezzella:2024tkf,Ovgun:2025bol}, as have (numerical) solutions for halos around stationary objects \cite{Fernandes:2025osu}. Although such constructions are recent and thus less-examined than the vacuum case, the consensus is clear: astrophysical environments can directly affect our current understanding of BH dynamics and GW astrophysics in the strong-field regime \cite{Cardoso:2022whc,Konoplya:2021ube,Xavier:2023exm,Mollicone:2024lxy,Spieksma:2024voy,Gliorio:2025cbh,Datta:2025ruh}.

Light rays \cite{wam98,treu10} or GWs \cite{Takahashi:2003ix,suv22} originating from a background source that pass by a compact object can be lensed. In cases with small-enough impact parameter, details of the lensed fluxes can provide not only tests of GR in the strong-field regime \cite{jain10,ezq2020,Chan:2025wgz} but also, in principle, reveal properties of the lens itself \cite{gond22,Massey:2010hh,Ellis:2010,Moustakas:2009na}. For instance, the first detection of an \emph{isolated} stellar-mass BH was achieved via the astrometric lensing of a field star lying in the Galactic bulge \cite{OGLE:2022gdj}, and light from the furthest-known star (Earendel; at a redshift of $z \approx 6.2$) was rendered measurable thanks to its emissions being magnified by factors up to $\sim 10^{4}$ due to its positioning relative to the galaxy cluster WHL0137--08 \cite{welch22}. 

Motivated by these recent discoveries and the other astrophysical elements described above, we revisit various aspects of lensing but for BHs embedded in halo environments. While lensing by dark matter halos has been extensively explored in the literature (see, e.g., Ref. \cite{mun16}), the use of an exact solution to Einstein's equations with a well-motivated matter halo \cite{Hernquist:1990be,Cardoso:2021wlq} has not. Nevertheless, the current literature uses (post-)Newtonian (PN) approximations of the leading and next-to-leading order terms to emulate the effects of the environment. Here, we also perform a comparison of deflection angles between fully-relativistic and PN configurations and find that only highly compact configurations have a relative difference $>1\%$ (but always $\sim10\%$), while in the the cases of comparatively low density that we studied extensively, a PN analysis more than suffices.

This paper is organized as follows. The exact solution we consider is described in Section \ref{sec:spacetime}, with a general overview of lensing theory provided in Section \ref{sec:lensing}. Results regarding the deflection angles (Sec. \ref{sec:deflection}) and amplification/bifurcation of images (Sec. \ref{sec:ampfactors}) produced by a BH surrounded by a matter halo configuration relative to the classical Schwarzschild and PN predictions are explored in Section \ref{sec:results}. These findings are applied in Section \ref{sec:astro} to consider an application with respect to ``echoes''. Conclusions are given in Section \ref{sec:conclusions}. Throughout most of this paper, geometric units with $G=c=1$ are used.

%%%%%%%%%%%%%%%%%%%%%%%%%%%%%%%%%%%%%%%%%%%%%%%%%%%%%%%%%%%%
\section{Static and spherically-symmetric black holes embedded in matter halos} \label{sec:spacetime}
%%%%%%%%%%%%%%%%%%%%%%%%%%%%%%%%%%%%%%%%%%%%%%%%%%%%%%%%%%%%

An exact, general-relativistic solution that describes a static and spherically-symmetric BH surrounded by a matter halo was proposed recently in Ref. \cite{Cardoso:2021wlq}. The construction of the environment follows the Einstein cluster scheme \cite{Einstein:1939ms} that assumes a collective of gravitating masses following all possible spherical geodesics, leading to an anisotropic fluid description, with zero radial pressure and non-zero tangential pressure $P_t$, such that
\begin{equation}
    T^\mu_\nu=\text{diag}(-\rho,0,P_t,P_t).
\end{equation}
Here, $\rho$ is the energy density of the matter profile. In Ref.~\cite{Cardoso:2021wlq}, the Hernquist profile was used, that describes dark matter halos around galaxies and bulges \cite{Hernquist:1990be}
\begin{equation}\label{Hernquist}
	\rho(r)=\frac{M a_0}{2\pi r (r+a_0)^3},
\end{equation}
where $M$ is the mass of the halo and $a_0$ is the length scale of the halo that designates the radius that contains $1/4$ of the configuration's mass, after which the density profile drops off as $\sim r^{-4}$. In the following we refer to $M/a_0$ as the compactness of the matter halo. The assumption of spherical symmetry, and the choice of a Hernquist-like matter distribution leads to the radially-dependent mass profile of the form
\begin{equation} \label{massprofile}
	m(r)=M_\text{BH}+\frac{M r^2}{(a_0+r)^2}\left(1-\frac{2M_\text{BH}}{r}\right)^2,
\end{equation}
that includes the BH mass, $M_\text{BH}$, in addition to the Hernquist contribution. In Schwarzschild-like coordinates $\{t,r,\theta,\phi\}$, the above eventually lead to the exact solution to the Einstein equations with line element 
\begin{equation}\label{metric}
	ds^2=-f(r)dt^2+\frac{dr^2}{1-2m(r)/r}+r^2 d\Omega^2,
\end{equation}
for 
\begin{subequations}
\begin{align}
	f(r)&=\left(1-\frac{2 M_\text{BH}}{r}\right)e^\Upsilon,\\
	\Upsilon&=-\pi\sqrt{\frac{M}{\xi}}+2\sqrt{\frac{M}{\xi}}\arctan\left[\frac{r+a_0-M}{\sqrt{M\xi}}\right],\\
	\xi&=2a_0-M+4 M_\text{BH},
\end{align}\label{eq:g_tt}
\end{subequations}
provided that $P_{t} = \rho/2$ (for any mass function).

The metric satisfies all expected limits. For instance, at small scales, equation \eqref{metric} describes a BH of mass $M_\text{BH}$, while at large scales the metric asymptotes to a spacetime with a Hernquist density profile \eqref{Hernquist}, described by the halo mass $M$ of length scale $a_0$. An event horizon resides at $r=r_{\rm Sch}=2 M_\text{BH}$, while at $r=0$ the spacetime has a curvature singularity. The configurations has Arnowitt-Deser-Misner (ADM) mass equal to the linear sum $M_\textrm{ADM}=M_\text{BH}+M$. 

To replicate galactic-scale dark matter halos, the hierarchical inequality $M_\text{BH}\ll M\ll a_0$ must hold. Furthermore, the compactness of the halo must satisfy $M/a_0\lesssim 10^{-4}$ \cite{Navarro:1995iw}, in order for the metric \eqref{metric} to describe supermassive BHs at the galactic core of the configuration, where the galaxy is surrounded by a Hernquist dark matter halo. Nevertheless, the mass and length scale of the halo are free parameters of the metric so it may even describe BHs within compact clouds of matter, as long as $M< 2(a_0+r_{\rm Sch})$ to avoid the appearance of pathological curvature singularities at
$r = M - a_0 \pm \sqrt{M^2 - 2\,M\, a_0 - 4\,M\,M_\textrm{BH}}$ \cite{Cardoso:2021wlq}.

The light ring, $r=r_\textrm{LR}$, resides at roots of the equation $r=3m(r)$ \cite{Cardoso:2008bp}. To leading order for any $\rho$, the light ring is displaced from its vacuum position as \cite{Cardoso:2021wlq}
\begin{equation} \label{light_ring_position}
    r_\textrm{LR}\simeq 3 M_\textrm{BH}\left[1+\frac{M \,M_\textrm{BH}}{a^2_0}+\mathcal{O}\left(a^{-3}_0\right)\right].
\end{equation}
The associated angular frequency, $\Omega_{\rm LR}$, and Lyapunov exponent, $\Lambda_{\rm LR}$ (instability timescale of photons at the light ring), are shifted to \cite{Cardoso:2021wlq}
\begin{align}
    M_{\rm BH}\Omega_{\rm LR}&\sim \frac{1}{3\sqrt{3}}\left[1-\frac{M}{a_0}+\frac{M(M+18M_{\rm BH})}{6a_0^2}+\mathcal{O}\left(a^{-3}_0\right)\right],\label{eq:lr_redshift}\\
M_{\rm BH}\Lambda_{\rm LR}&\sim \frac{1}{3\sqrt{3}}\left[1-\frac{M}{a_0}+\frac{M^2}{6a_0^2}+\mathcal{O}\left(a^{-3}_0\right)\right],\label{eq:lr_lyapunov}
\end{align}
respectively. From the above, it is clear that photons are gravitationally red-shifted according to the compactness of the halo, $M/a_0$. The critical impact parameter $b_{\rm crit}$ for capture of high-frequency photons or GWs is \cite{Cardoso:2021wlq}
\begin{equation}\label{critical_impact_parameter}
 b_{\rm crit}\simeq3\sqrt{3}M_{\rm BH}\left[1+\frac{M}{a_0}+\frac{M(5M-18M_{\rm BH})}{6a_0^2}+\mathcal{O}\left(a^{-3}_0\right)\right].  
\end{equation}
The fact that the relevant parameters introduced above shift relative to the Schwarzschild case indicates that impinging radiation will be deflected in a way that depends sensitively on the compactness of the halo.

%%%%%%%%%%%%%%%%%%%%%%%%%%%%%%%%%%%%%%%%%%%%
\subsection{Mass profile} \label{sec:massprofile}
%%%%%%%%%%%%%%%%%%%%%%%%%%%%%%%%%%%%%%%%%%%%

In this paper, we consider a range of compactness values, $M/a_0$, spanning $10^{-4}$ and $10^{-1}$. Such scales may delimit the range of plausible extrema for halos at moderate-to-high redshifts. For example, Ref.~\cite{somer18} used data from the Cosmic Assembly Near-infrared Deep Extragalactic Legacy (CANDELS) and Galaxy and Mass Assembly (GAMA) surveys to estimate the relationship between galaxy and halo size out to redshifts $\sim 3$. Their figure 2 for instance illustrates that a typical compactness ranges between $10^{-3} \lesssim M/a_0 \lesssim 10^{-6}$ -- softer than that considered here (noting their $r_{e}$ denotes a \emph{half-mass} radius not quarter as for the Hernquist $a_0$). However, massive galaxies at high redshift are thought to form via dissipative processes involving mergers or disc instabilities, leading to more compact configurations \cite{porter14}. In particular, galaxies typically become increasingly gas-poor as a function of time, and if massive galaxies seen closer to the present epoch form via sequential gas-poor galaxy mergers they will naturally be less compact \cite{somer18}. Larger values of $M/a_0$ may thus be expected for galaxies at higher redshifts \cite{zolo15,well15}, for which it can be difficult to estimate the compactness directly. We caution the reader therefore that while there is little, direct observational evidence that compactness values can reach $\lesssim 10^{-1}$ in realistic systems, we consider a wide parameter range of the metric in order to explore the effects of halos on lensing observables.

%%%%%%%%%%%%%%%%%%%%%%%%%%%%%%%%%%%%%%%%%%%%
\section{Lensing theory} \label{sec:lensing}
%%%%%%%%%%%%%%%%%%%%%%%%%%%%%%%%%%%%%%%%%%%%

Gravitational lensing arises when the path of light from a distant source is bent by the gravitational potential of an intervening mass distribution (the lens). The theoretical description of lensing involves several key physical scales: in observational terms, lensing angles are typically measured in arcseconds, a scale that is well-matched to the small angular separations observed in many lensing systems. The characteristic scale of image separation is set by the \textit{Einstein radius}, which is defined as the angular radius at which light from a perfectly aligned background source is focused into a ring, known as an \emph{Einstein ring}. The Einstein radius, \( \theta_E \), is given by (in radians)
\begin{equation}
\theta_E = \sqrt{\frac{4GM_\textrm{L}}{c^2} \frac{D_\textrm{LS}}{D_\textrm{OL} D_\textrm{OS}}},
\end{equation}
where \( M_{\rm L} \) is the mass of the lens,
\( D_{\rm OL} \) is the angular diameter distance between the observer and the lens,
\( D_{\rm OS} \) is the distance between the observer and the source, and \( D_{\rm LS} \) is the distance between the lens and the source. The distances \( D_{\rm OL} \), \( D_{\rm OS} \), and \( D_{\rm LS} \) are calculated under the assumption of a given cosmological model (e.g., a \(\Lambda\)CDM universe) and depend on the redshifts of the observer, lens, and source. The Einstein radius thus encapsulates the geometry of the lensing system and the mass of the lens. For a lens with mass \( M_{\rm L} \sim 10^{12} M_\odot \) at redshift \( z_{\rm L} \sim 0.5 \) and a source at \( z_{\rm S} \sim 2.0 \), for example, the Einstein radius is typically on the order of \( \sim 1'' \).

We introduce the lens-plane coordinates $\boldsymbol{x}$, representing a (dimensionless) \emph{angular} position in the lens plane. Explicitly, one usually measures angles in units of the Einstein radius \(\theta_E\), so that \( x=\theta/\theta_E\). Working within the thin-lens approximation where one of the spatial dimensions is collapsed by assuming the mass distribution of the lens is infinitely thin in some plane, we have that $x$ is two dimensional. Mathematically, this implies that the diffraction integral describing the complex amplification factor (including phase shifts) can be written as \cite{suv22,gb23},
\begin{equation} \label{eq:diffint}
    F(w,y)=\frac{w}{2\pi i}\int d^2x\;\exp\bigl[\,i\,w\,T(x,y)\bigr]\,, 
\end{equation}
where \(w=\omega(1+z_{\rm L} (D_{\rm L}D_{\rm S}/D_{\rm LS})\theta_E^2\) is the dimensionless frequency (in terms of the physical angular frequency $\omega$), $\boldsymbol{y}$ is the (dimensionless) angular position of the source in the source plane, and
\begin{equation}
    T(x,y)=\frac12\,|x-y|^2-\psi(x)+\phi_m(y)
\end{equation}
is the Fermat potential. The symbol \(\psi( x)\) denotes the lensing potential, given by the two‐dimensional Poisson integral of the convergence \(\kappa(x)\)
\begin{equation}
      \psi (x)
      = \frac{1}{\pi}\int\mathrm{d}^2x' 
      \kappa ( x')
      \ln\bigl|x -  x'\bigr| ,
\end{equation}
where the projected surface density is defined as 
\begin{equation} \label{eq:sigma}
    \Sigma(x) \equiv \int dr \rho(x,r),
\end{equation}
such that the convergence \(\kappa(x)\) is the ratio 
\begin{align}
\kappa(x) \equiv \frac{\Sigma(x)}{\Sigma_{\text{cr}}}, \quad\quad \Sigma_{\text{cr}} \equiv \frac{c^2}{4\pi G}\frac{D_{\rm OS}}{D_{\rm OL} D_{\rm LS}}.
\end{align}
Physically, \(\nabla_{\!x}\,\psi(x)\) equals the deflection angle \(\alpha( x)\). The constant \(\phi_m\) is chosen so that the Fermat potential has a convenient zero‐point -- it may be set to zero without loss of generality.

%%%%%%%%%%%%%%%%%%%%%%%%%%%%%%%%%%%%%%%%%%%%%%%%%%
\subsection{Geometric optics}\label{sec:geooptics}
%%%%%%%%%%%%%%%%%%%%%%%%%%%%%%%%%%%%%%%%%%%%%%%%%%

In practice, evaluating the integral \eqref{eq:diffint} is challenging as it oscillates out to infinity (though see Refs. \cite{feld19,suv22} for some mathematical methods). However, lensing can be adequately described within the \textit{geometric optics} framework when the wavelength of radiation, $\lambda \equiv 2 \pi c /\omega$, is much smaller than the characteristic scale of the lens, $r_{\text{lens}}$, i.e., $\lambda \ll r_{\text{lens}} \sim 2 G M_{\rm L}/c^2$. Given the magnitude of $M_{\rm L}$ in many astrophysical applications of lensing by BH+halo systems, this approximation is reasonable. In this limit, radiation propagates along null geodesics of the metric \eqref{metric}, and wave effects (diffraction, interference) are negligible.   Calculating the integral \eqref{eq:diffint} in the geometric-optics limit reduces to finding the stationary points \(\{x_j\}\) (i.e., images) satisfying the lens equation
\begin{equation} \label{eq:lenseqn}
    \nabla_xT(x_j,y)=0,
\end{equation}
such that the amplification factor reduces to the sum over individual images,
\begin{equation}
F_{\rm GO}(w,y)
=\sum_j\sqrt{|\mu(x_j)|}
\;\exp\bigl[i\,w\,T(x_j,y)\bigr]\,,
\end{equation}
where 
\begin{equation} \label{eq:magfac}
\mu=\left[\det\left(\frac{\partial \boldsymbol{y}}{\partial \boldsymbol{x}}\right)\right]^{-1}
\end{equation}
is the magnification of each image. 

Our spacetime consists of a general-relativistic, static and spherically symmetric BH of mass $M_\text{BH}$ embedded in a Hernquist dark matter halo of mass $M$ and scale radius $a_0$, with mass profile \eqref{massprofile}. This configuration introduces two main curvature scales. Near the BH horizon ($r \sim 2 M_\text{BH}$), where the spacetime is approximately Schwarzschild, the curvature radius is $\mathcal{R}_\text{BH} \sim M_\text{BH}$. At large radii ($r \gg M_\text{BH}$), where the halo dominates, the Hernquist profile leads to curvature scale $\mathcal{R}_\text{halo} \sim a_0$. For realistic galactic halos, the inequalities $M_\text{BH} \ll M \ll a_0$ and $M/a_0 \lesssim 10^{-4}$ must hold, leading to a hierarchy:
\begin{equation}
    \mathcal{R}_\text{BH} \sim M_\text{BH} \ll \mathcal{R}_\text{halo} \sim a_0.
\end{equation}
The smallest scale $\mathcal{R}_\text{BH}$ sets the most stringent requirement on the wavelength $\lambda$ for geometric optics to hold. We refer the reader to \cite{wam98} for a review. 

%%%%%%%%%%%%%%%%%%%%%%%%%%%%%%%%%%%%%%%%%%
\subsection{Strong vs. weak-field lensing} \label{sec:optics}
%%%%%%%%%%%%%%%%%%%%%%%%%%%%%%%%%%%%%%%%%%

\begin{figure*}
    \centering
    \includegraphics[width=\textwidth]{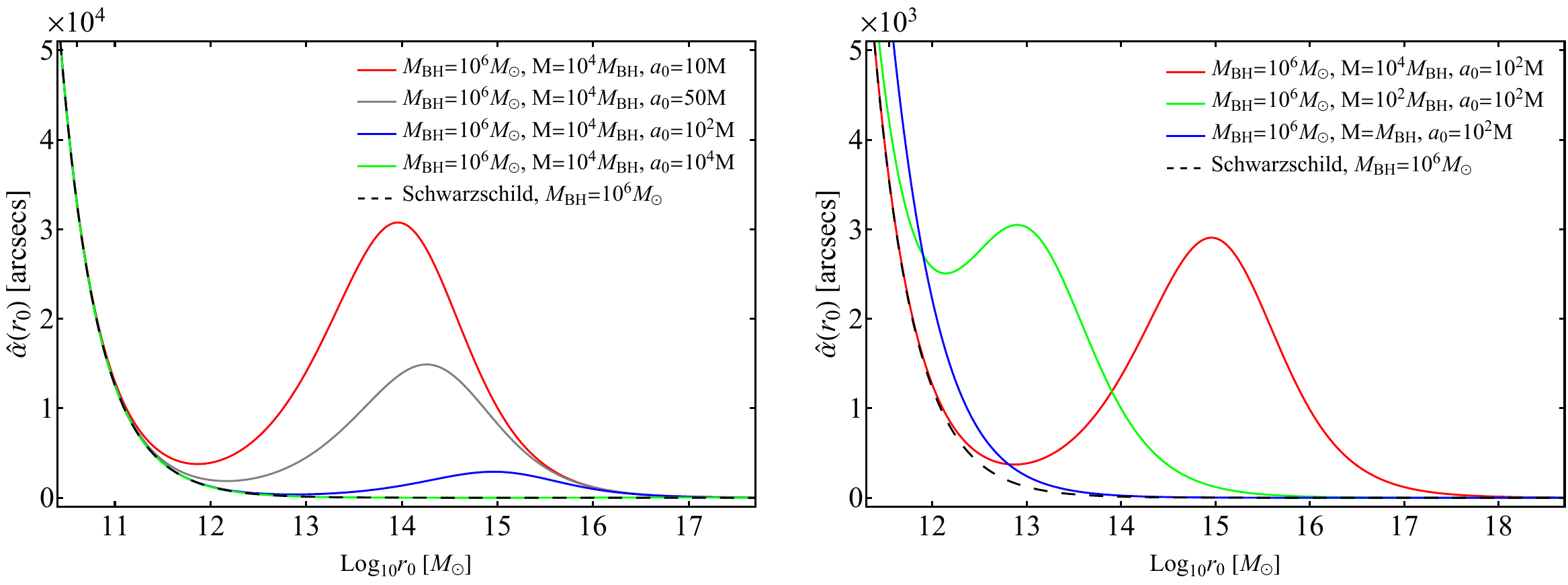} 
    % FIGURE 1
\caption{Total bending angle $\hat{\alpha}(r_0)$ (arcsec) versus photon periastron, $r_0$.
\emph{Left:} Compactness $M/a_0$ is varied by changing $a_0$ at fixed $(M_{\rm BH},M)=(10^6 M_\odot,\,10^4 M_{\rm BH})$; curves show $a_0/M\in\{10,50,10^2,10^4\}$ alongside the Schwarzschild baseline (see figure legends). Increasing compactness boosts bending across all $r_0$ and shifts the strong-lensing upturn outward (larger $r_0$), reflecting the outward shift of the light ring and larger capture impact parameter. \emph{Right:} Halo mass is varied instead at fixed $a_0/M=10^2$; curves show $M/M_{\rm BH}\in\{1,10^2,10^4\}$ plus Schwarzschild. Larger $M$ similarly enhances bending and advances the transition into the strong-lensing regime; overall amplitude is chiefly controlled by the compactness $M/a_0$.}
    \label{deflections_plot}
\end{figure*}

The deflection angle $\hat{\alpha}$ for a photon with periastron $r_0$ is computed via the exact integral formulation for spherically symmetric spacetimes, i.e.,
\begin{equation}\label{deflection_equation}
\hat{\alpha}(r_0) = 2\int_{r_0}^\infty \frac{dr}{r^2\sqrt{\dfrac{1}{b^2} - \dfrac{f(r)}{r^2}}} - \pi, \quad b = \frac{r_0}{\sqrt{f(r_0)}},
\end{equation}
where $b$ denotes the impact parameter (cf. expression \ref{critical_impact_parameter}). This approach remains valid across both weak and strong gravitational regimes as it incorporates the full nonlinear metric structure without approximation. 

Weak lensing arises when the projected surface density \(\Sigma(\theta)\) from expression \eqref{eq:sigma}
is everywhere below the critical value \(\Sigma_{\rm cr}\)
so that \(\kappa(\theta) \ll 1\) and no multiple images form in equation \eqref{eq:lenseqn}.  In this regime one may linearize the lens mapping,
\begin{equation}
    \beta_i\approx \theta_i-\partial_i\psi(\theta),
\end{equation}
where $\beta$ is the (two‑dimensional) angular position of the source on the sky in the absence of any deflection. The components $\beta_i$ label the orthogonal angular coordinates in the source plane for $\theta$ being the observed angular position of an image of that source, again with components $\theta_i$. In practice, one measures $\theta_i$ directly on the sky and $\psi(\theta)$ is the (projected) lensing potential, which encodes the integrated gravitational effect of the deflector. Strong lensing occurs when \(\kappa(\theta)\gtrsim1\) in some regions, producing multiple images separated by angles \(\Delta\theta\sim\theta_E\).  For a point‐mass lens, multiple images form at
\begin{equation}
\theta_\pm=\tfrac12(\beta\pm\sqrt{\beta^2+4})\,\theta_E. 
\end{equation}
For the weak-field regime ($r_0 \gg r_{\text{Sch}} + a_0$), the Hernquist potential dominates and yields $\hat{\alpha} \sim 4M_{\text{halo}}/r_0$ asymptotically, consistent with linearized gravity predictions. In the strong-field regime ($r_0 \lesssim 10\, r_{\text{Sch}}$), the integral captures nonlinear effects near the light ring where deflection angles diverge logarithmically. The implementation, detailed in Appendix \ref{sec:numerical}, handles metric singularities at $r = 2M_{\text{BH}}$ through the exponential suppression in $f(r)$ and utilizes adaptive quadrature for numerical stability across $10^5 \leq r_0/M_\odot \leq 10^{12}$. This coverage ensures quantitative comparison of halo-induced deviations from pure Schwarzschild lensing at all observable scales of interest.

Suppose that $O$ and $O_{\rm Schw}$ denote lensing observables associated with a case with a halo and the Schwarzschild BH, respectively, with the same $M_{\rm BH}$. By defining the residuals
\begin{equation}
\begin{aligned}
    \delta O(x) &\equiv O(x)-O_{\rm Schw}(x), \\
\end{aligned}
\end{equation}
we say that a ``{bump}'' is present if $\delta O$ has a localized positive maximum at $x=x_b$, i.e.\ $\partial_x \delta O|_{x_b}=0$, $\partial^2_x \delta O|_{x_b}<0$, and $\delta O(x_b)>0$. Here, $x$ denotes either $\log(r_0/M_{\rm BH})$ if considering the deflection angle or $\beta$ if considering the magnification.

The bump \emph{does not} require a local maximum in the matter density, nor is it unique to the relativistic case. Instead, it arises from the interplay of:
(i) a global upward shift of the weak-field tail from the extended Hernquist envelope, which adds a $\sim 1/r_0$ contribution to $\hat\alpha$ and raises $\mu(\beta)$ at fixed $\beta$; and
(ii) a halo-induced outward displacement of the photon sphere and increase in the capture threshold, which advances and sharpens the strong-lensing upturn. Mechanism (ii) follows from the analytic shifts
$r_{\rm LR}\simeq 3M_{\rm BH}\!\left[1+ (M M_{\rm BH}/a_0^2)+\cdots\right]$,
a reduced $\Omega_{\rm LR}$ and $\Lambda_{\rm LR}$, and an increased $b_{\rm crit}$ (Eqs.~(\ref{light_ring_position})–(\ref{critical_impact_parameter})), which collectively move the onset of the logarithmic divergence in $\hat\alpha$ to larger $r_0$ compared to vacuum. The superposition of (i) and (ii) naturally produces a localized excess in the residual $\delta O$ centered near the transition from the halo-dominated regime to the (shifted) near–light-ring regime.

%%%%%%%%%%%%%%%%%%%%%%%%%%%%%%%%%%%%%
\section{Results} \label{sec:results}
%%%%%%%%%%%%%%%%%%%%%%%%%%%%%%%%%%%%%

We now move to the main goal of this work: understanding how the presence of a halo may influence lensing observables.

%%%%%%%%%%%%%%%%%%%%%%%%%%%%%%%%%
\subsection{Deflection angles}\label{sec:deflection}
%%%%%%%%%%%%%%%%%%%%%%%%%%%%%%%%%

We compute the total (bending) deflection angle $\hat\alpha(r_0)$ from expression \eqref{deflection_equation} in the BH+halo spacetime, and present it in arcseconds for direct comparison across models. This integral keeps both the weak- and strong-field regimes and reduces to the standard Schwarzschild result in the vacuum limit, so any departures in the curves isolate the halo imprint. 

In Fig. \ref{deflections_plot} on the left, we vary the compactness $M/a_0$ at fixed $(M_{\rm BH},M)$. Increasing compactness---i.e., shrinking $a_0$ at fixed halo mass---pushes the entire $\hat\alpha(r_0)$ profile upward at all radii. Physically, the exponential redshift factor $f(r)=(1-2M_{\rm BH}/r)\,e^{\Upsilon}$ deepens the effective potential and increases the total enclosed mass inside a given $r_0$, so rays experience larger bending than in vacuum along the whole trajectory. In particular, the near–light-ring upturn is enhanced and its onset shifts to slightly larger $r_0$ (i.e., farther from the horizon) because the halo moves the light ring outward by a fractional amount $\Delta r_{\rm LR}/r_{\rm LR}\sim(M_{\rm BH}M/a_0^2)$, cf.\ Eq. \eqref{light_ring_position}, and increases the critical capture parameter $b_{\rm crit}$; see equation \eqref{critical_impact_parameter}. The net effect is an earlier (in $r_0$) and stronger transition into the strong-lensing regime, where multiple images form and signals can be effectively received more than once by the observer (see also Sec.~\ref{sec:astro}).
For the most compact halos (smallest $a_0$), this manifests as order-of-magnitude enhancements relative to Schwarzschild across the full domain; as $a_0$ grows (compactness decreases) the curves converge monotonically to the vacuum baseline, leaving only percent-level, yet systematic, offsets at astrophysically realistic $M/a_0\lesssim10^{-4}$. 

\begin{figure} 
    \centering     \includegraphics[width=0.485\textwidth]{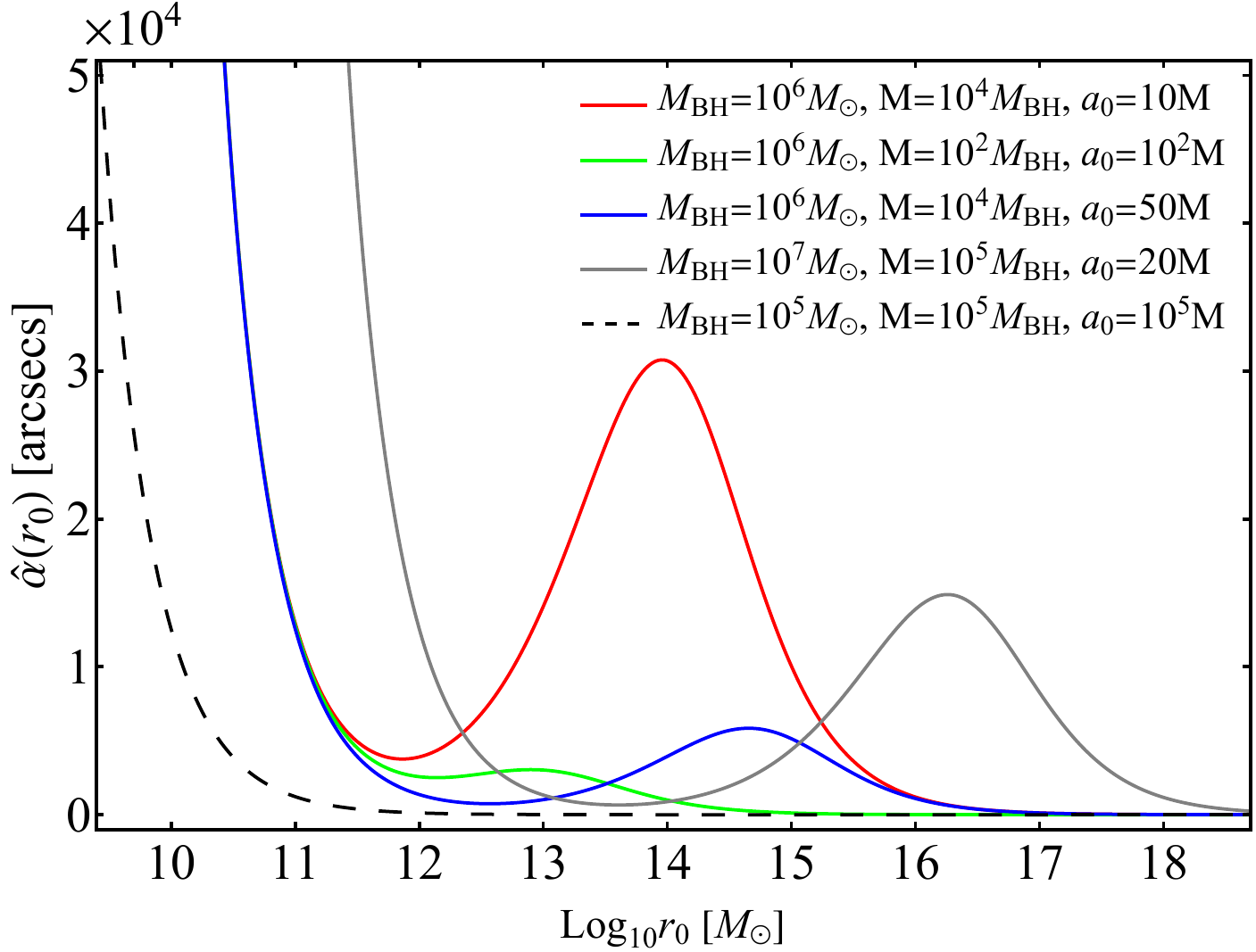} 
\caption{Similar to Fig.~\ref{deflections_plot}, though instead showing combined parameter variations and degeneracies. Changing $M_{\rm BH}$ primarily rescales the horizontal axis (via $r_{\rm Sch}\!\propto\!M_{\rm BH}$), whereas compactness $M/a_0$ sets the overall amplitude and sharpness of the near-light-ring upturn. The figure highlights a degeneracy between central mass and environmental compactness if only $\hat{\alpha}(r_0)$ is used, suggesting joint constraints from weak- and strong-field portions of the curves.}\label{fig:var_all}
\end{figure}

The right panel of Fig. \ref{deflections_plot} shows results where instead the halo mass $M$ is varied at fixed $M/a_0$. Increasing $M$ has a qualitatively similar impact: stronger overall bending and a shift of the strong-lensing turn-up to larger $r_0$. The reason is the same redshift (or ``deeper potential'') mechanism quantified by Eqs. \eqref{light_ring_position}--\eqref{critical_impact_parameter}: larger $M$ at fixed $a_0$ pushes $r_{\rm LR}$ outward and raises $b_{\rm crit}$, so more trajectories that would have produced weak bending in vacuum now undergo large deflections.
The figure makes explicit that the amplitude of $\hat\alpha$ is primarily controlled by the compactness $M/a_0$ rather than by $M$ or $a_0$ in isolation, which is clear from the grouping of curves by $M/a_0$. 

The behavior observed in the right panel of Figure~1, where the deflection angle $\hat{\alpha}(r_0)$ is larger for smaller halo masses $M$ at small radii, arises from the interplay between the halo's mass distribution and the location of the photon sphere. In the strong-field regime ($r_0 \lesssim 10\,r_{\text{Sch}}$), the deflection angle is highly sensitive to the position of the photon sphere $r_{\text{LR}}$, which is shifted outward relative to the Schwarzschild case due to the presence of the halo (Eq.~(\ref{light_ring_position})). For a fixed halo scale radius $a_0$, increasing the halo mass $M$ leads to a larger compactness $M/a_0$, which in turn shifts $r_{\text{LR}}$ further outward. As a result, for a given small $r_0$, a model with a larger $M$ has a photon sphere that is farther from $r_0$, thereby reducing the deflection angle compared to a model with a smaller $M$, where the photon sphere is closer to $r_0$ and the lensing is stronger.
This effect is consistent with the analytic expressions for $r_{\text{LR}}$ and $b_{\text{crit}}$ (Eqs.~(\ref{light_ring_position}) and (\ref{critical_impact_parameter})), which show that both increase with $M$ at fixed $a_0$. Therefore, the flip in the magnitude relation at small $r_0$ reflects the competition between the overall mass increase (which tends to enhance lensing) and the outward shift of the photon sphere (which delays the strong-lensing upturn). At larger $r_0$, the halo's enclosed mass dominates, and the expected monotonic behavior with $M$ is restored.

In the weak-field ($r_0\!\gg\! r_{\rm Sch}+a_0$) the Hernquist envelope supplies an additional $1/r_0$ tail to the bending, so the outer branches of all halo curves lie above Schwarzschild with a nearly parallel slope; differences here are set by the enclosed halo mass at the impact radius and scale with $M/a_0$. As $r_0$ decreases, nonlinear terms become important and the curves steepen toward the logarithmic divergence at the light ring; the halo shifts this onset to larger $r_0$, expanding the parameter space that yields multiple images and large time delays. 

When $(M_{\rm BH},M,a_0)$ are varied simultaneously in  Fig. \ref{fig:var_all}, the families of curves demonstrate two clean pieces of physics: (i) changing $M_{\rm BH}$ rescales the overall length scale (shifting the horizontal axis via $r_{\rm Sch}\propto M_{\rm BH}$), while (ii) changing the compactness $M/a_0$ sets the vertical offset and the sharpness/onset of the strong-lensing rise. Models with larger $M_{\rm BH}$ but smaller $M/a_0$ can partially mimic the near-field behavior of lighter BHs embedded in denser halos, underlining a degeneracy between central mass and environmental compactness if only $\hat\alpha(r_0)$ is used. Conversely, simultaneous fits to the far-field slope (weak lensing; halo-dominated) and the location of the near-field upturn (set by $r_{\rm LR}$ and $b_{\rm crit}$) can break this degeneracy. 

The upward displacement of $\hat\alpha(r_0)$ at all $r_0$ implies that halo-dressed lenses systematically over-bend relative to vacuum, biasing any Schwarzschild-based inference toward a heavier central mass if environmental effects are ignored. Because the weak-field tail and the strong-field onset respond differently to $M_{\rm BH}$ and $M/a_0$, joint constraints using wide-separation images (or astrometric deflections) together with relativistic-image observables offer a route to measure the halo compactness directly from deflection-angle data. 

%%%%%%%%%%%%%%%%%%%%%%%%%%%%%%%%%%%%%%%%%%%%
\subsection{Newtonian, post-Newtonian and full metric comparison} \label{sec:pn}
%%%%%%%%%%%%%%%%%%%%%%%%%%%%%%%%%%%%%%%%%%%%
\begin{figure*}
    \centering
    \includegraphics[width=\textwidth]{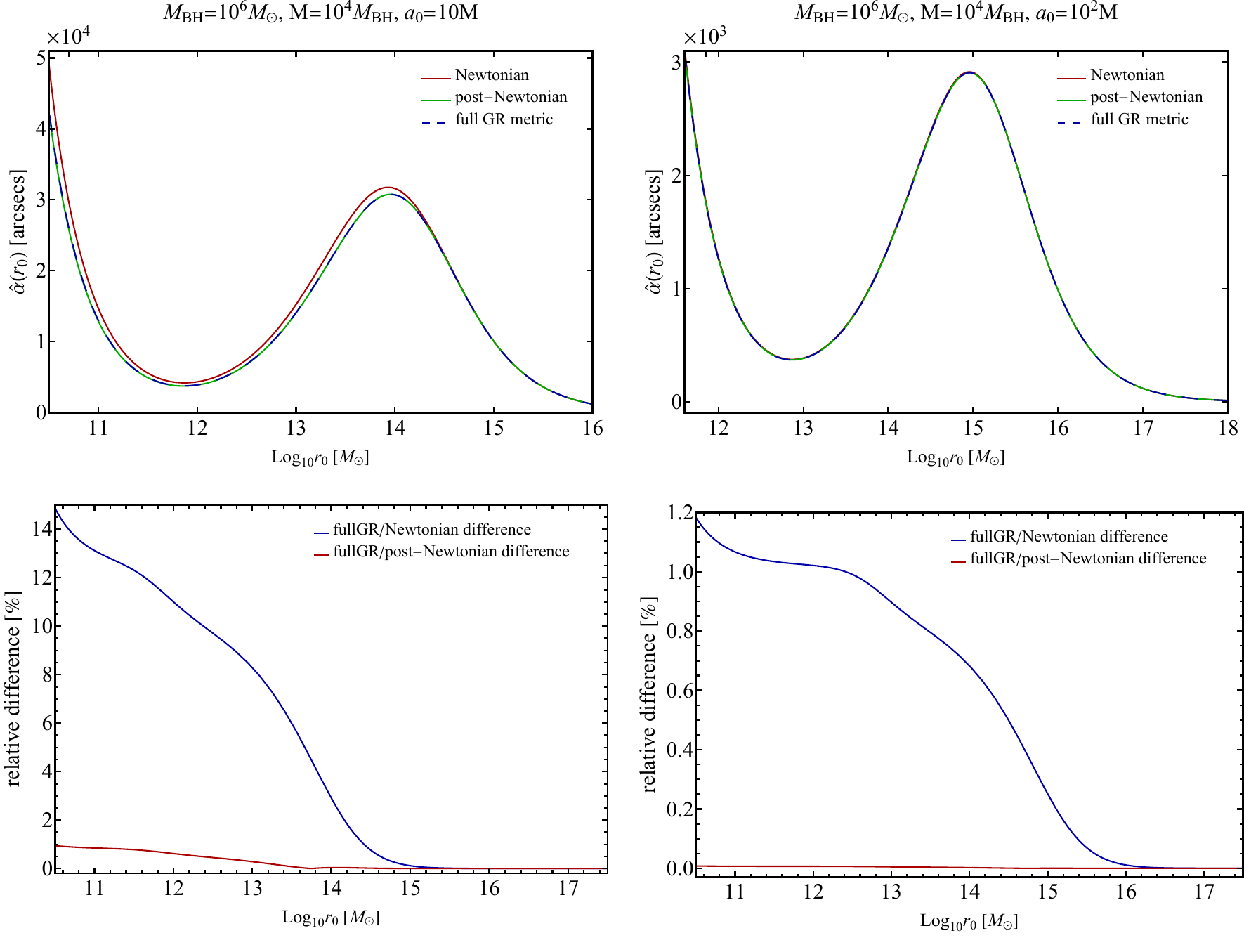} 
\caption{Total bending angle $\hat{\alpha}(r_0)$ (arcsec) versus photon periastron, $r_0$. Here, we use the full metric \eqref{eq:g_tt}, the Newtonian ($1/c^2$) and the PN expansion of $f(r)$, through the Eq. \eqref{eq:PN_expansion}.
\emph{Top panel:} Compactness is set to $M/a_0=10^{-1}$ (left), with fixed $(M_{\rm BH},M)=(10^6 M_\odot,\,10^4 M_{\rm BH})$; curves show the total bending angle of the Newtonian and PN metric with respect to the full solution. On the right panel the comparison is the same but with $M/a_0=10^{-2}$.
\emph{Bottom panel:} The relative difference of total bending angles of the full GR solution and the Newtonian, PN approximations of the metric. It is obvious that smaller compactness leads to an order of magnitude less error.}
\label{fig:deflections_PN}
\end{figure*}

Throughout this work, we investigate lensing predictions within a fully-relativistic framework. Depending on the parameter ranges under consideration however, it may be that a PN framework can accurately capture the salient features. The simplest way in which a direct comparison can be made is to expand the metric potentials in powers\footnote{We caution the reader that this naive expansion does not fully capture what is often meant by PN ``metrics'', as one usually uses isotropic coordinates and particular gauges so that the spatial slices are conformally flat; see, e.g., expression (1) in Ref.~\cite{suv18}.} of $c^{-2}$ and recompute the physical observables, such as the deflection angle \eqref{deflection_equation}.

Restoring units of $c$ temporarily we find from expression \eqref{eq:g_tt}, for instance, that
\begin{equation}
\begin{aligned}\label{eq:PN_expansion}
 f(r) =& 1+ -\frac{\frac{2 M}{a_{0}+r}+\frac{2 M_{\rm BH}}{r}}{c^2} \\
 &+ \frac{4 M \left(a_{0} M+\frac{3 M_{\rm BH} (a_{0}+r)^2}{r}\right)}{3 c^4
   (a_{0}+r)^3} + \mathcal{O}(c^{-6}).
   \end{aligned}
\end{equation} 
This expansion encapsulates how the total deflection angle behaves with respect to the the compactness of the halo at PN order. In Fig. \ref{fig:deflections_PN} it is shown that the compactness $M/a_0$ plays a key role the validity of PN vs. relativistic results. Configurations with $M/a_0\gtrsim10^{-1}$ show distinctive relative percentage errors of $\mathcal{O}(10)$ in the Newtonian approximation (see left in Fig. \ref{fig:deflections_PN}), while smaller compactness reduces the error significantly (right column). The error reduces for a PN treatment. This lines up with the expected behavior: when the halo compactness is sufficiently small, a PN approximation of particle dynamics is quite accurate, with the relative difference between PN and exact solution deflection angles being less than $1\%$. On the other hand, when the compactness is considered to be large, i.e., $M/a_0=10^{-1}$, then a Newtonian treatment is inadequate at the strong-field regime, though a PN approximation is still of order $\mathcal{O}(10^0)\%$.

\begin{table}[]
    \centering
\scalebox{1.15}{
    \begin{tabular}{c|c|c}
         \hline\hline
         $\beta$ (arcseconds) & $\theta$ (arcseconds) & $\mu$ \\
         \hline
        $10^{-4}$ & 1.1576 & 5786.99 \\
         ($10^{-4}$)  & (1.1576) & (5788.21) \\\hline
          $10^{-2}$ & 1.1626 & 58.37 \\
          ($10^{-2}$) & (1.1626) & (58.38) \\\hline
          1 & 1.7668 & 1.23 \\
          (1) & (1.7668) & (1.23) \\\hline
          2 & 2.5322 & 1.04 \\
          (2) & (2.5296) & (1.05) \\
        \hline\hline
    \end{tabular}
}
\caption{Positions of relativistic images ($\theta$) and magnification factors ($\mu$) produced by the non-vacuum metric in Eq. \eqref{metric} as a function of angular impact parameter ($\beta$), at its vacuum limit, with $M_\textrm{BH}=2.8\times 10^6 M_\odot$, $M=10^{-6}M_\textrm{BH}$ and $a_0=10^2M$. For reference, the resulting values are compared with those obtained for a galactic Schwarzschild BH with $M_\textrm{BH}=2.8\times 10^6 M_\odot$ in Table II of Ref.~\cite{Virbhadra:1999nm} in parentheses.}\label{tab:table}
\end{table}

\begin{figure*} 
    \centering
    \includegraphics[width=\textwidth]{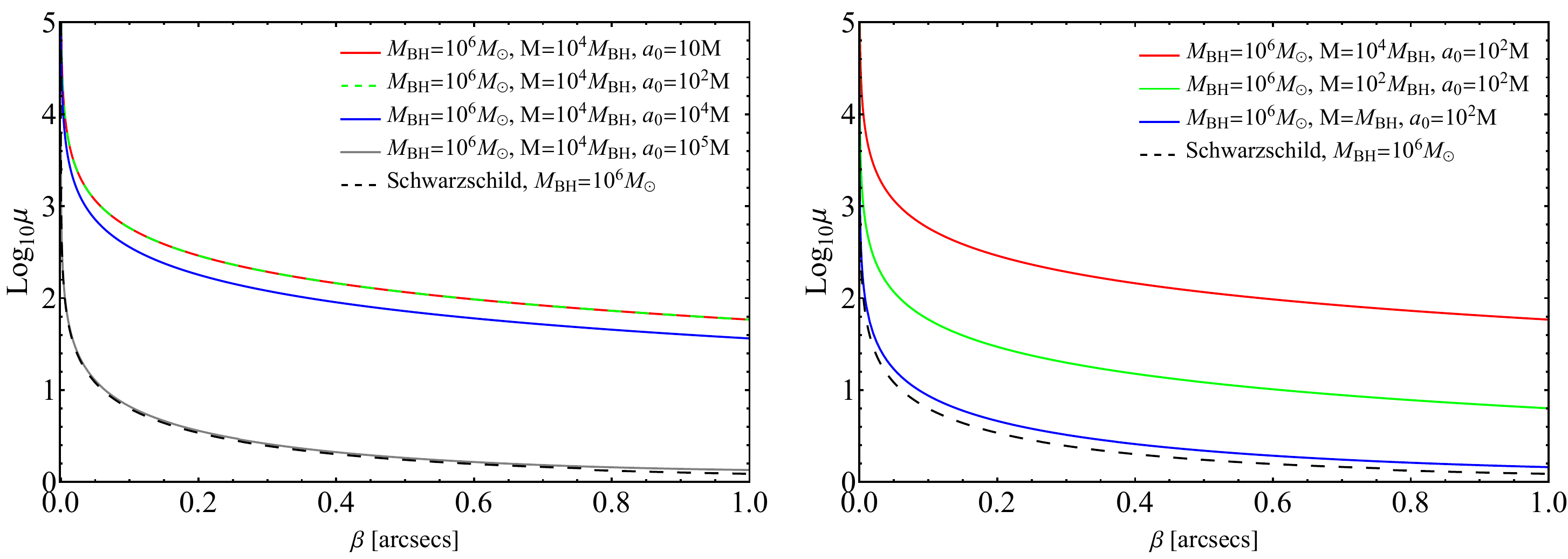} 
\caption{Amplification factors, $\mu$, as a function of source offset, $\beta$.
\emph{Left:} Compactness $M/a_0$ varied at fixed $(M_{\rm BH},M)=(10^6 M_\odot,\,10^4 M_{\rm BH})$ with $a_0/M\in\{10,10^2,10^4,10^5\}$, with the Schwarzschild case shown for comparison (see figure legends). \emph{Right:} Halo mass varied at fixed $a_0/M=10^2$ with $M/M_{\rm BH}\in\{1,10^2,10^4\}$ plus Schwarzschild.
Denser or more massive halos systematically raise $\mu$ at fixed $\beta$ and extend the domain of strong magnification to larger angular offsets; in the diffuse/low-mass limit the curves converge to the vacuum result.}
    \label{Magnifications_plot}
\end{figure*}

%%%%%%%%%%%%%%%%%%%%%%%%%%%%%%%%%%
\subsection{Magnification factors}
\label{sec:ampfactors}
%%%%%%%%%%%%%%%%%%%%%%%%%%%%%%%%%%

Figure \ref{Magnifications_plot} shows the total geometric–optics magnification $\mu(\beta)$ as a function of the (dimensionless) angular impact parameter. In our setup the magnification is mathematically given by equation \eqref{eq:magfac}, which expresses the fact that it encodes how the lens mapping stretches a given solid angle on the sky. In the left panel we vary the halo compactness $M/a_0$ at fixed $(M_{\rm BH},M)$, while in the right panel we vary $M$ at fixed $a_0$; in both cases the Schwarzschild vacuum curve is shown for reference.

Relative to the Schwarzschild case, every halo-dressed model sits above the vacuum curve across the full range of $\beta$. Physically, the Hernquist envelope redshifts photon dynamics and deepens the effective potential, $f(r)=(1-2M_{\rm BH}/r)e^{\Upsilon}$, pushing the light ring outward and raising the critical capture parameter $b_{\rm crit}$ [cf.\ Eqs. \eqref{light_ring_position}--\eqref{critical_impact_parameter}]. As a result, rays that would have undergone modest deflection in vacuum are bent more strongly, so the same source offset $\beta$ yields a larger $\mu$. This manifests as a nearly uniform upward shift of $\mu(\beta)$ together with an extension of the strong-lensing regime to larger angular separations: high-magnification events remain possible even when the source is not (almost) perfectly aligned.
\\
The left panel makes clear that increasing $M/a_0$ (shrinking $a_0$ at fixed $M$) boosts $\mu$ at all $\beta$ and delays the falloff of the curves; the right panel shows that raising $M$ at fixed $a_0$ produces the same qualitative trend. Taken together, the families group primarily by the ratio $M/a_0$ rather than by $M$ or $a_0$ alone: compactness controls both the vertical increase/offset of $\mu$ and the $\beta$ at which the strong-lensing upturn sets in. In the near-alignment limit the expected $\mu\!\propto\! \beta^{-1}$ divergence remains visible, but the halo shifts the whole relation upward; at larger $\beta$ the outer branches run roughly parallel to the Schwarzschild tail, indicating similar asymptotics with a halo-set offset (set by the enclosed mass at the relevant impact radii).

Two practical consequences follow. First, if one interprets $\mu(\beta)$ with a vacuum model, the over-bending induced by the halo biases lens-mass inferences high. Second, because strong magnifications persist to larger $\beta$ in compact halos, the cross-section for highly magnified events increases: for example, Fig. \ref{Magnifications_plot} (left) shows that even $\beta\!\sim\!1''$ can deliver $\mu\!\sim\!10^2$ once $a_0\!\lesssim\!10^4 M$ (keeping $M$ fixed). In joint analyses this environmental degeneracy can be broken by combining the shape of $\mu(\beta)$ with deflection-angle information and/or time delays $\Delta t(\beta)$ (see Sec. \ref{sec:astro}), since the weak-field tail and the strong-field onset respond differently to $M_{\rm BH}$ and $M/a_0$.

In the vacuum limit of our metric (taking $M\!\to\!0$ at fixed $M_{\rm BH}$) the image positions and magnifications in Table \ref{tab:table} coincide with the standard Schwarzschild values, providing a useful sanity check of the amplification pipeline used to generate Fig.~\ref{Magnifications_plot}.

%%%%%%%%%%%%%%%%%%%%%%%%%%%%%%%%%%%%%%%%%%%%%%%%%%
\section{An application: echoes} \label{sec:astro}
%%%%%%%%%%%%%%%%%%%%%%%%%%%%%%%%%%%%%%%%%%%%%%%%%%

The potential barrier associated with gravitational perturbations of an exotic compact object can exhibit a secondary bump or ``wall'' far outside of the horizon such that individual modes may become temporarily trapped \cite{Cardoso:2016rao,Cardoso:2016oxy,Price:2017cjr,vp17,Cardoso:2019rvt}. As familiar from studies of such potentials in quantum-mechanical settings, a wavepacket that enters this intermediate region can bounce between the two walls and gradually leak out in a manner that can be mathematically described by the reflection and transmission (RT) coefficients. Any given quasinormal mode (QNM) \cite{ks99,Berti:2009kk,Destounis:2023ruj,Berti:2025hly} may therefore be fragmented into multiple pulse trains, with those beyond the initial being called \emph{echoes} \cite{Ferrari:2000sr}. 

Echo templates have been considered in \cite{mark17}, where it was shown that a variety of RT coefficients could be expected depending on the system of interest. On the other hand, lensing may also produce distinct images of various amplitudes and time-delays,
\begin{equation}\label{eq:timelag}
\Delta t \approx \frac{4 G M_{\rm L}}{c^3} \left [ \frac y 2 \sqrt{y^2+4} + \log\left(\frac{\sqrt{y^2+4}+y}{\sqrt{y^2+4}-y}\right)\right].
\end{equation}
Depending on the impact parameter and/or scale of the halo, echoes may not be the smoking gun for exotic objects or modified gravity \cite{Moffat:2014aja, Zhao:2022gxl,Capozziello:2012zj,Vlachos:2021weq,Chatzifotis:2021pak,Kouniatalis:2024gnr} they are often considered to be. For instance, in $\lesssim 10\%$ of cases where a GW impinges on a supermassive BH to within $\sim 200$ Schwarzschild radii, an echo of amplitude may $\sim 0.1$ may be received \cite{gond22} due to multi-image formation (see also \cite{Conklin:2017lwb}). Quantifying such effects is therefore important to eventually establish whether a hypothetically-detected echo \cite{Uchikata:2019frs,Uchikata:2023zcu} has an astrophysical origin (i.e., a lens) or is associated with the source \cite{Abedi:2016hgu,Maggio:2019zyv,Abedi:2020ujo,Testa:2018bzd,Maggio:2020jml,Chakraborty:2022zlq}.

In considering cases with a halo, the probability of lensing could increase significantly depending on the compactness, $a_{0}$. As can be seen from the left-hand panel of Fig. \ref{deflections_plot}, for instance, even angular impact parameters of $\sim 1$~arcsecond could lead to significant magnifications ($\sim 100$) if $a_{0} \lesssim 10^{4}M$. While a statistical analysis lies beyond the scope of this paper, comparing such a case with that of Schwarszschild, we may expect an echo of similar amplitude to that reported in Ref.~\cite{gond22} but for cases reaching out to $\sim 10^{3}$ Schwarzschild radii. Based on expression \eqref{eq:timelag}, if the lens mass is held fixed but the angular impact parameter increases by a factor $\sim 5$, an increase of a factor $\sim$~few in $\Delta t$ would be expected for $\beta \sim \mathcal{O}(0.1)$. In general therefore, cases with a halo would likely produced greater delays between echos compared to BHs in vacuum. By contrast, the functional form of the magnification curves seen in Fig. \ref{Magnifications_plot} show a similar falloff for cases with halos, indicating a comparable amplitude just with a skew towards greater impact parameters. Comparing the blue and dashed curves in the right-hand panel as a conservative case, at the same impact parameter we may expect a factor $\sim 2$ increase in the echo amplitude. As expected physically since the lens is overall heavier, predictions for echos of \emph{astrophysical origin} for cases with halos involve a skew towards larger amplitude and greater time delays.

An interesting possibility in this respect is that of catching an echo without the main signal. It is possible that, since the deflection angle is large even at many Schwarzschild radii (see Fig. \ref{fig:var_all}), an image is diverted into the path of the observer by the lens while the main signal is not. This would significantly shift the parameter interpretation from GW template matching to QNMs from a remnant BH, neutron star, or other compact object. The Kerr QNM frequencies scale inversely with mass \cite{ks99}, and thus receiving an echo of relative magnitude $\sim 0.2$ -- without the main signal -- would skew the mass inference of the object by a factor $\sim 5$. If halos are very extended in Nature, such considerations should be accounted for in GW pipelines to avoid the possibility of unphysical inferences on compact-object nature and tests of GR more generally (see, e.g., Refs.~\cite{s19,ksds22,skk24} for discussions in a variety of contexts). Significant amplifications could also render a GW signal detectable from an astrophysical source that would otherwise have low signal-to-noise ratio (e.g. from giant flares \cite{zink12}), similar to how Earendel was detected \cite{welch22}.

%%%%%%%%%%%%%%%%%%%%%%%%%%%%%%%%%%%%%%%%%%%%
\section{Conclusions}\label{sec:conclusions}
%%%%%%%%%%%%%%%%%%%%%%%%%%%%%%%%%%%%%%%%%%%%

In this paper, we present an analysis of how lensing by BHs in astrophysical environments manifests via the deflection angles (Sec. \ref{sec:deflection}) and amplification factors (Sec. \ref{sec:ampfactors}). Although a thorough comparison with realistic astrophysical data cannot be facilitated with the simple model developed here, some qualitative features stand out. The main results are highlighted in Figures \ref{fig:var_all} and \ref{Magnifications_plot}, showing how a smooth Hernquist profile will generally lead to an overdense ``bump'' in the respective profiles at radii far from the horizon, depending on the compactness of the halo, $M/a_{0}$. A comparative analysis of the resulting deflection angles between PN approximations, and the metric configuration itself supports the validity of PN schemes to mimic environmental effects, such as light bending, but only in cases where the compactness is sufficiently small. As emphasized earlier, the fully relativistic treatment mainly provides a
precision confirmation rather than revealing a qualitatively new
effect as the bump appears even at the Newtonian level. Focusing on ``echoes'' (Sec. \ref{sec:astro}) -- traditionally associated with beyond-GR compact objects and quantum effects at the horizon scale that generalize BHs in some appropriate sense \cite{vp17,Cardoso:2019rvt,Maggio:2021ans,Maggio:2023fwy} -- we argue that these extended features could increase the probability of multiple images being received by the observer as the strong-lensing regime is maintained at larger radii than the case of a singular, compact BH. 

In reality, the matter profile of halos are unlikely to be smooth but rather ``lumpy''. Astrophysical material will typically condense into individual objects (e.g., stars or ``dark stars'') which will introduce some coarse graining into the profile. By using a different mass density profile to that of expression \eqref{Hernquist}, via a sum of delta functions or otherwise, the impact of this could be studied. In the case of star clusters, methods based on Picard-Lefschetz integrals have been used to solve the full diffraction integral \eqref{eq:diffint} \cite{feld19,suv22}, necessary since the individual lenses making up the overall macrolens may not satisfy the geometric optics limit. When the dimensionless frequency, \(w=\omega(1+z_{\rm L})(D_{\rm L}D_{\rm S}/D_{\rm LS})\theta_E^2\), is finite and small, diffractive corrections become important, which leads to frequency‐dependent modulations of the observed waveform(s) and breaks the simple superposition of ray‐optics images. This could lead to more complex phenomenology than that seen in the geometric optics limit, which is the main focus of this paper.

Aside from wave-optical extensions, we have used a static background here. Using the stationary generalization of the spacetime metric from \cite{Fernandes:2025osu} it would be relatively straightforward to extend the analysis here to see how rotation impacts on the results (see also Ref. \cite{feld25}). From an astrophysical perspective, there are many other systems where lensing is thought to play a role. One interesting example concerns the elusive fast radio bursts (FRBs), often stipulated to be associated with quaking magnetars \cite{wang18,sk19}. Many of these $\sim$~mHz radio transients are seen to repeat at various intervals. Those without any periodicity are candidates for lensing: it could be that the observer receives a (modulated) signal more than once \cite{mun16}, rather than repetition being intrinsic to the source. Since the formalism we have presented is independent of whether the radiation is composed of light or GWs, our results could be applied in that context to assess probabilities of lensing by extended halos around BHs (see, e.g., Ref.~\cite{li25}).

%%%%%%%%%%%%%%%%%%%%%%%
\begin{acknowledgments}
We thank the anonymous referee for providing helpful comments, which improved the quality of this manuscript.
A.G.S. was supported by the Conselleria d'Educaci{\'o}, Cultura, Universitats i Ocupaci{\'o} de la Generalitat Valenciana through Prometeo Project CIPROM/2022/13 during the early stages of this work.
K.D. is indebted to the Physics Department of the School of Applied Mathematical and Physical Sciences of National Technical University of Athens for the hospitality provided during the early stages of this work.
K.D. acknowledges financial support provided by FCT – Fundação para a Ciência e a Tecnologia, I.P., under the Scientific Employment Stimulus – Individual Call – Grant No. 2023.07417.CEECIND/CP2830/CT0008.
This project has received funding from the European Union’s Horizon MSCA-2022 research and innovation programme “Einstein Waves” under grant agreement No. 101131233.
\end{acknowledgments}
%%%%%%%%%%%%%%%%%%%%%%%

\appendix

%%%%%%%%%%%%%%%%%%%%%%%%%%%%%%%%%%%%%%%%%%%%%%%%%
\section{Numerical methods} \label{sec:numerical}
%%%%%%%%%%%%%%%%%%%%%%%%%%%%%%%%%%%%%%%%%%%%%%%%%

We work in geometrized units ($G=c=1$) and define the BH$+$halo metric via equation \eqref{massprofile}. The Schwarzschild baseline is obtained by replacing $f(r),a(r)\to 1-2M_{\rm BH}/r$. For a photon with periastron $r_0$, the impact parameter is $b=r_0/\sqrt{f(r_0)}$. The (reduced) deflection integral, equation \eqref{deflection_equation}, is computed using the substitution $u=r/r_0$ so that the domain is $u\in[1,\infty)$ and the endpoint $u=1$ (the turning point) remains integrable. 

The BH$+$halo Einstein angle $\theta_E$ is found by solving the lens equation at perfect alignment, $F(\theta)=\theta-\alpha(\theta)=0$,
via bracketing and bisection using the \emph{exact} $\alpha(\theta)$ (no interpolant dependence for $\theta_E$). Around $\theta_E$ we build a nonuniform $\theta$–grid (dense near $\theta_E$, logarithmic to larger angles), evaluate $\alpha(\theta)$ exactly at the grid points, and construct a piecewise–linear interpolant. The derivative $\alpha'(\theta)$ entering $\mu$ is evaluated from local quadratic least–squares fits to three neighboring points for numerical stability. Beyond the largest grid angle we attach the analytic weak–field tail
\begin{equation*}
\alpha(\theta)\simeq \frac{4(M_{\rm BH}+M)}{D_{\rm L}\,\theta},
\quad
\alpha'(\theta)\simeq -\frac{4(M_{\rm BH}+M)}{D_{\rm L}\,\theta^2},
\end{equation*}
ensuring the correct $1/\theta$ falloff (no constant clamping). The primary–image magnification is computed as
\[
\mu(\theta)=\Big[(1-\alpha(\theta)/\theta)\,(1-\alpha'(\theta))\Big]^{-1}.
\]
For specified offsets $\beta$, we solve $\theta-\alpha(|\theta|)=\beta$ (again by bracketing/bisection) and evaluate $\mu$ at the solution.

%%%%%%%%%%%%%%%%%%%%%
\bibliography{biblio}
%%%%%%%%%%%%%%%%%%%%%

\end{document}